\documentclass[aps,prb,twocolumn,groupedaddress]{revtex4}
\usepackage{txfonts}
\usepackage[dvips]{epsfig}

\begin{document}
\title{Precise measurements of electron and hole g-factors of single quantum dots by using nuclear field}
\author{R.\ Kaji}
	\affiliation{Department of Applied Physics, Hokkaido University, Sapporo 060-8628, Japan}
	
\author{S.\ Adachi}
    \email[]{adachi-s@eng.hokudai.ac.jp}    
    \altaffiliation[Also at ]{CREST, Japan Science and Technology Agency, Kawaguchi 332-0012, Japan}
	\affiliation{Department of Applied Physics, Hokkaido University, Sapporo 060-8628, Japan}
		
\author{H.\ Sasakura}
	\altaffiliation[Also at ]{CREST, Japan Science and Technology Agency, Kawaguchi 332-0012, Japan}
	\affiliation{Department of Applied Physics, Hokkaido University, Sapporo 060-8628, Japan}
	
\author{S.\ Muto}
    \altaffiliation[Also at ]{CREST, Japan Science and Technology Agency, Kawaguchi 332-0012, Japan}
	\affiliation{Department of Applied Physics, Hokkaido University, Sapporo 060-8628, Japan}
	
\date{\today}

\begin{abstract}
We demonstrated the cancellation of the external magnetic field by the nuclear field at one edge of the nuclear polarization bistability in single InAlAs quantum dots. The cancellation for the electron Zeeman splitting gives the precise value of the hole g-factor.  
By combining with the exciton g-factor that is obtained from the Zeeman splitting for linearly polarized excitation, the magnitude and sign of the electron and hole g-factors in the growth direction are evaluated.
\end{abstract}
\maketitle

Semiconductor quantum dots (QDs) exhibit a variety of confinement-related optical and electronic properties useful for opto-electronic device applications such as QD lasers and detectors. In particular, broad efforts are currently underway to develop new techniques for controlling spin degrees of freedom in QDs for quantum information processing. A key quantity for the spin manipulation is a g-factor, which is a coefficient connecting magnetic dipole moments with the spin degrees of freedom.
Therefore, the knowledge of electron and hole g-factors and their control are important. 
For example, zero electron g-factor is required to convert the photon qubit into the electron spin qubit~\cite{{Kosaka01}} while the system with a large g-factor is preferable
for controlling spin qubit in terms of the energy selectivity.
The g-factors of self-assembled QDs have been obtained by optical measurements and transport measurements. Generally, the electron g-factor is deduced from transport measurements
while the exciton g-factor, which is the sum of an electron and a hole g-factors, is deduced from optical measurements. In the optical measurements, since the photoluminescence (PL) is generated by the annihilation of an electron and a hole, it is usually difficult to independently obtain an electron or a hole g-factor of QDs. In addition, sensitivity of the g-factors to the spatial confinement has been predicted by theoretical studies and partly confirmed in the experiments~\cite{Nakaoka04}.
The obtained values of the g-factors are much different from bulk ones possibly due to size quantization, strain, and other effects, however those effects are difficult to evaluate nondestructively and noncontactly for individual QD. Therefore, the direct probing method of the electron or hole g-factor is required for individual QD target.

In this study, we demonstrate the precise measurements of electron and hole g-factors in single InAlAs QDs by using the optically induced nuclear field. The measurement principle is based on the fact that nuclear field is effective only on electrons and can compensate the external magnetic field. 
We first show that the nuclear field exactly cancels the external magnetic field  at one edge of nuclear bistability.  Recently, we proposed to use the cancellation of external and nuclear field for photon-spin qubit conversion to dispense with the zero g-factor engineering~\cite{Muto05}. The cancellation at the bistability indicates that the condition for the qubit conversion is automatically realized there.
Then, the measured Zeeman splitting corresponds to that of not an exciton but a hole, and therefore the hole g-factor is obtained precisely. In order to raise the precision of the obtained value, we measured the hole g-factor at some different excitation powers. Together with the exciton g-factor that is evaluated by the linearly polarized excitation, the electron g-factor is obtained.

We use the self-assembled In$_{0.75}$Al$_{0.25}$As/Al$_{0.3}$Ga$_{0.7}$As QDs which were grown by molecular beam epitaxy on a (100) GaAs substrate~\cite{Sasakura04}. 
Single QDs are studied by using a standard micro-PL ($\mu$-PL) setup with a microscope objective lens.
To limit the observable QD number, small mesa structure with a typical top lateral diameter of $\sim$150 nm was fabricated~\cite{Yokoi05}. The $\mu$-PL measurements were performed at 5 K under the magnetic field $B_{\rm z}$ up to 5 T in Faraday geometry. For excitation, cw Ti:sapphire laser was employed.
The excitation light polarization was controlled precisely by the polarization selective optics and  waveplates. The $\mu$-PL from QDs was collected by the same microscope objective and was dispersed by a triple-grating spectrometer, and was detected with a liquid-nitrogen cooled Si-CCD camera.
The system resolution was $\sim$12 $\mu$eV and the spectral resolution that determines the resonant peak energies was found to be less than 5 $\mu$eV by the spectral fitting. The typical exposure time of the CCD camera was 1 sec. to obtain one spectrum with a high signal-to-noise ratio. 

\begin{figure}[t]
  \begin{center}
    \includegraphics[width=240pt]{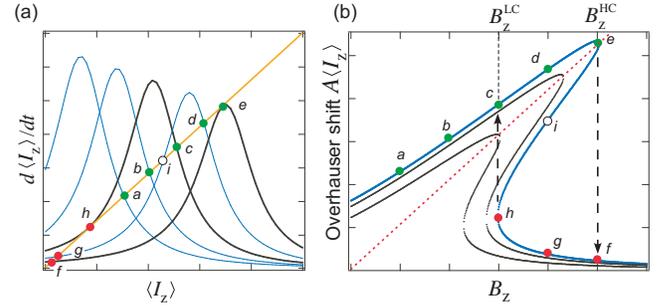}
\caption{Graphical representation of the dependence of OHS on $B_{\rm z}$ for $\sigma^-$ excitation. (a) The steady-state $\langle {I_{\rm z} } \rangle$ of Eq.~\ref{eq1} is determined by the balance between the Lorentzian-shaped polarization term and the depolarization term (a straight line). The thick curves represent the polarization terms at the low and high critical $B_{\rm z}$ ($B_{\rm N}^{\rm LC}$ and $B_{\rm N}^{\rm HC}$) for the bistable behavior. (b) The resultant OHS shows the bistable response on $B_{\rm z}$. The curves for different excitation powers are also depicted. The curve with a larger excitation power has the larger $B_{\rm N}^{\rm HC}$. Note that the top of the curves is always on the line of $|g_{\rm z}^{\rm e}| \mu_{\rm B} B_{\rm z}$ (a dotted line).} \label{Fig1}
    \end{center}
\end{figure}
Since the g-factor measurement in this study is based on the cancellation of $B_{\rm z}$ for an electron in a target QD by the optically induced nuclear field $B_{\rm N}$, we begin with the explanation of the response of $B_{\rm N}$ on $B_{\rm z}$. 
Figure~\ref{Fig1}(a) shows the $B_{\rm z}$-dependence of a nuclear spin polarization term (Lorentzian-shaped function) and depolarization term (straight line) of the following rate equation that represents the nuclear spin dynamics~\cite{Abragam61,OptOrientation};
\begin{equation}
\frac{{d\left\langle {I_{\rm z} } \right\rangle }}{{dt}} = \frac{1}{{T_{\rm{NF}} }}\left[ Q \left( {\left\langle {S_{\rm z} } \right\rangle  - S_{0} } \right)-{\left\langle {I_{\rm z} } \right\rangle} \right] - \frac{1}{{T_{\rm{ND}} }}\left\langle {I_{\rm z} } \right\rangle, \label{eq1}
\end{equation}
where $\langle {I_{\rm z} } \rangle$ and $\langle {S_{\rm z} } \rangle$ are the averaged nuclear and electron spin polarizations, $S_{0}$ is the thermal electron spin polarization, $1/T_{\rm{NF}}$ and $1/T_{\rm{ND}}$ are the nuclear spin polarization and depolarization rates, and 
$Q \ (=\left[{I\left( {I + 1} \right)}\right]/\left[{S(S + 1)}\right])$ is a momentum conversion coefficient from electron spin to nuclear spin system in the spin flip-flop process. 
Based on the general form of the spin-flip process in the precessional decoherence type~\cite{OptOrientation}, the spin transfer rate $1/T_{\rm{NF}}$ is given as follows by assuming the uniform electron wavefunction in a QD~\cite{Braun06,Tartakovskii07,Maletinsky07};
\begin{equation}
\frac{1}{T_{\rm{NF}} }= \left[\frac{n_{\rm e} \tau_{\rm c}^2}{\tau_{\rm R}} \left(\frac{A }{N \hbar} \right)^2\right] \bigg/ \left[1+ \left(\frac{\tau_{\rm c}}{\hbar} \right)^2 \left\{g_{\rm z}^{\rm e} \mu_{\rm B} (B_{\rm z} \pm B_{\rm N}) \right\}^2 \right],
\label{eq2}
\end{equation}
where $A$, $N$, $n_{\rm e}$, and $\tau_{\rm R}$ are the hyperfine coupling constant, number of nuclei, electron density in the QD, and lifetime of an electron, respectively. $\tau_{\rm c}$ is the correlation time of the coupled electron-nuclei system with the broadening $\hbar/\tau_{\rm c}$. The magnitude of $B_{\rm N}$ is related to the steady state $\left\langle I_{\rm z} \right\rangle$ as $B_{\rm N}=A \left\langle I_{\rm z} \right\rangle / |g_{\rm z}^{\rm e}| \mu_{\rm B}$ and the direction can be selected to parallel or anti-parallel to $B_{\rm z}$ by the choice of the helicity of circular excitation.
According to Eq.~\ref{eq2}, the compensation of $B_{\rm z}$ by $B_{\rm N}$ reduces the electron Zeeman splitting i.e. $g_{\rm z}^{\rm e} \mu_{\rm B} (B_{\rm z} - B_{\rm N})$ and results in the increase of  $1/T_{\rm{NF}}$.
From this simple model, the coupled electron-nuclear spin system shows a static hysteresis loop in the relation between the Overhauser shift (OHS=$A\langle I_{\rm z} \rangle$) and three variable parameters: $n_{\rm e}$ ($\propto$ excitation power), $\langle {S_{\rm z} } \rangle$($\propto$ excitation polarization) and $B_{\rm z}$. 
By some groups and us, the hysteretic behavior of OHS in QDs are recently observed on the excitation power~\cite{Braun06,Tartakovskii07,Kaji07}, on the excitation polarization~\cite{Kaji07}, and on $B_{\rm z}$~\cite{Maletinsky07}. 

\begin{figure}[t]
  \begin{center}
    \includegraphics[width=240pt]{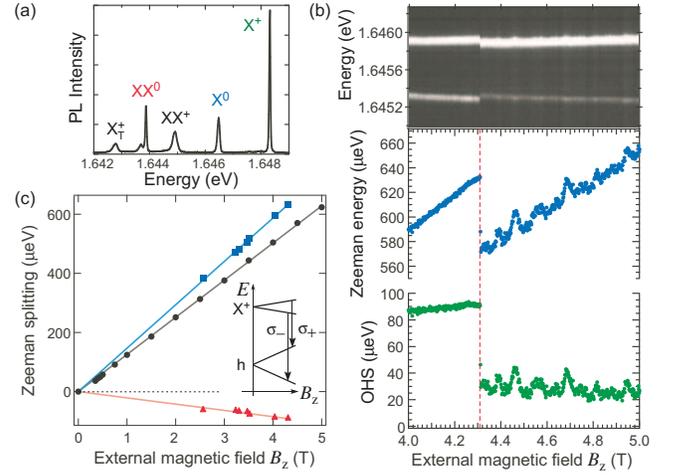}
  \end{center}
  \caption{(a) PL spectra of a single QD at 0 T (5 K). (b) Contour plot of the Zeeman-split PL lines of $X^+$ for $\sigma^-$-excitation as a function of $B_{\rm z}$ (upper panel). The $B_{\rm z}$-dependence of Zeeman splitting energy (middle panel). The $B_{\rm z}$-dependence of OHS (bottom panel). The abrupt decrease of the splitting energy at $\sim$4.3 T is observed clearly in all figures. (c) The observed Zeeman splitting energies of hole (squares), electron (triangles), and exciton (circles) are plotted as a function of $B_{\rm z}$. Inset shows the transitions of $X^+$ PL.} 
  \label{Fig2}
\end{figure}
The steady state $\langle {I_{\rm z}} \rangle$ of Eq.~\ref{eq1} are expressed graphically by the intersecting points of both polarization and depolarization terms, as shown in Fig.~\ref{Fig1}(a), and the trajectory of the steady state $\langle {I_{\rm z} } \rangle$ is depicted as a function of $B_{\rm z}$ in Fig.~\ref{Fig1}(b). Both figures are depicted for $B_{\rm N}$ antiparallel to $B_{\rm z}$, which is realized by $\sigma_-$ excitation for InAlAs QDs.
The Lorentzian shaped polarization term moves right with increasing $B_{\rm z}$. 
At the low $B_{\rm z}$, the intersection is unique (e.g. point $a$) and then follows the straight line of the depolarization term with \textit{increasing} $B_{\rm z}$. Beyond the lower critical magnetic field $B_{\rm z}^{\rm LC}$, two new solutions (e.g. $g$ and $i$) appear, but the system still remains in the high $\langle {I_{\rm z}} \rangle$ state. Just beyond the higher critical magnetic field $B_{\rm z}^{\rm HC}$, this high $\langle {I_{\rm z}} \rangle$ state (point $e$) disappears and the $\langle {I_{\rm z}} \rangle$ jumps down to the low $\langle {I_{\rm z} } \rangle$ state (point $f$). For further increase in $B_{\rm z}$, the state stays on the lower branch. 
In the case of decreasing $B_{\rm z}$, the state on the lower branch remains at $B_{\rm z}^{\rm LC}$ where the low $\langle {I_{\rm z}} \rangle$ state $h$ disappears and the system has to return to the upper branch (point $c$). In the region between $B_{\rm z}^{\rm LC}$ and $B_{\rm z}^{\rm HC}$, there are two stable states (e.g. $d$ and $g$) and which one of them is realized depends on the history, i.e. on whether one comes from larger or smaller $B_{\rm z}$. The intermediate branch is unstable (e.g. point $i$). 
Such bistable and hysteretic behaviors of OHS on $B_{\rm z}$ were clearly observed in the low range of $B_{\rm z}$ ($\le$2 T) in single In(Ga)As QDs by Maletinsky et al~\cite{Maletinsky07}. 
Note that the following two points in this nonlinear behavior; \\
1. Perfect cancellation of $B_{\rm z}$ by $B_{\rm N}$ is achieved at $B_{\rm z}^{\rm HC}$ since the OHS at the time is the same as the electron Zeeman splitting $|g_{\rm z}^{\rm e}|\mu_{\rm B} B_{\rm z}$ (see the dotted line in Fig.~\ref{Fig1}(b)). \\
2. Non-zero OHS is predicted at $B_{\rm z}$=0. (This was observed experimentally~\cite{Eble06,Lai06}.)\\
From the above explanation, the hole g-factor $g_{\rm z}^{\rm h}$ can be obtained from the Zeeman splitting at $B_{\rm z}^{\rm HC}$, where $B_{\rm z} - B_{\rm N}=0$ is achieved and the electron Zeeman splitting is zero exactly. 

PL spectra of a target single InAlAs QD at 0 T is shown in Fig.~\ref{Fig2} (a). The spectra was obtained at the wetting layer (WL) excitation ($\sim$730 nm) by the depolarized light. The excitation gives rise to the transition between the WLs in conduction and valence bands.
The figure shows almost all emission lines from an isolated QD for the WL excitation with the moderate power.  
Through various measurements for assignments of PL spectra~\cite{Kumano06}, we conclude that the PL lines in the figure originate from the same single QD and can be attributed to $X^{+}_{T}$ (triplet state of positively charged exciton), $XX^{0}$ (neutral biexciton), $XX^{+}$ (positively charged biexciton), $X^{0}$ (neutral exciton), and $X^{+}$ (singlet state of positively charged exciton) from the low energy side. 
Hereafter, we focus on the $X^{+}$ PL because the PL is strongest in the case of WL excitation and $X^+$ PL has no fine structure due to no exchange interaction. 

Figure~\ref{Fig2}(b) shows a typical nonlinear behavior of the steady state $\langle {I_{\rm z}} \rangle$ around $B_{\rm z}^{\rm HC}$ for $\sigma_-$-excitation with a fixed excitation power; the contour plot of the Zeeman-split PL intensity (upper panel), Zeeman splitting energy (middle panel), and OHS (bottom panel) in the $B_{\rm z}$ range of $4-5$ T. In the upper panel, the high (low) energy line is $\sigma_-$($\sigma_+$)-polarized PL and the separation corresponds to the Zeeman splitting affected by $B_{\rm N}$, i.e. $g_{\rm z}^{\rm h} \mu_{\rm B} B_{\rm z}+g_{\rm z}^{\rm e} \mu_{\rm B} (B_{\rm z}-B_{\rm N} )$. Remember that $B_{\rm N}$ is effective only for electrons.
 
As shown in Fig~\ref{Fig1}(b), before the abrupt change the $B_{\rm N}$ overcompensates $B_{\rm z}$, i.e. $B_{\rm z}-B_{\rm N} < 0$. Increasing $B_{\rm z}$ under a constant excitation power, $B_{\rm z}-B_{\rm N}=0$ is achieved and then abrupt change of the Zeeman splitting occurs at $B_{\rm z}^{\rm HC}$=4.3091 T according to the aforementioned bistable nature.   
At this $B_{\rm z}$, the electron Zeeman splitting is zero and only the hole Zeeman splitting remains.
Then, from the Zeeman splitting, we can directly deduce the hole g-factor $g_{\rm z}^{\rm h}$. Note that the slopes of the Zeeman splitting, i.e. g-factors, before and after the abrupt decrease are clearly different due to the cancellation of the electron Zeeman splitting. 
From the bottom panel, the OHS decrease from $\sim$90 $\mu$eV to $\sim$30 $\mu$eV at $B_{\rm z}^{\rm HC}$. This decrease of OHS just corresponds to the drop from the point $e$ to the point $f$ in Fig.~\ref{Fig1}(b). 

The full width $2 \hbar/\tau_{\rm c}$ of the Lorentzian polarization term is estimated to be $\sim$30 $\mu$ eV since the coherence time was measured to be $\sim$45 ps in the temporal domain by the single-photon Fourier spectroscopy and in the spectral domain for this QD under the WL excitation~\cite{Adachi07}. Therefore, the point $e$ in Fig.~\ref{Fig1}(b) is very closely located on the peak of the narrow Lorentzian polarization term together with a long depolarization time (several ms) and the difference between $B_{\rm z}^{\rm HC}$ and $B_{\rm N}$ is negligible. 
After the abrupt decrease, the Zeeman splitting and OHS show the small fluctuation, which implies the existence of the excitation power fluctuation. 
Since $B_{\rm z}^{\rm HC}$ is a function of the excitation power as shown in Fig.~\ref{Fig1}(b),
if the power fluctuates positively, the corresponding changes of the OHS and Zeeman splitting are found  after the abrupt decrease. 
Therefore, by systematically changing the excitation power, we can observe the the abrupt change at different $B_{\rm z}^{\rm HC}$ in order to raise the precision of the deduced g-factors. The obtained values of the hole Zeeman splitting energy are plotted (squares) as a function of $B_{\rm z}^{\rm HC}$ in Fig.~\ref{Fig2}(c). We measured also the Zeeman splitting energies of $X^+$ PLs at the linearly polarized excitation (circles), which represents the difference of the transition energies as shown in the inset. From both data, we deduce the electron Zeeman splitting energy as a function of $B_{\rm z}$. Consequently, $g_{\rm z}^{\rm h}$ and  $g_{\rm z}^{\rm e}$ are evaluated as $+2.54 \pm 0.01$ and $-0.37 \pm 0.02$, respectively, by linear fitting. An electron g-factor in InAlAs QDs has the opposite sign of that in In(Ga)As QDs\cite{Bayer01,Bayer02}. The method used here is very powerful to obtain not only the magnitude but also the sign of electron and hole g-factors in a target single QD. This work provides valuable information on InAlAs/AlGaAs QDs which have been reported so far by only a few studies about the exciton g-factor~\cite{Hinzer01}. 

In summary, we proposed the method of individual evaluation of electron and hole g-factors of single InAlAs quantum dots by using the optically induced nuclear field. The cancellation of $B_{\rm z}$ by $B_{\rm N}$ for the electron Zeeman splitting gives the precise value of the hole g-factor.  
By combining it with the Zeeman splitting for linearly polarized excitation, the magnitude and sign of the electron and hole g-factors in the growth direction are obtained. The measurements by using the nuclear field gives not only a distinctive technique to obtain the g-factors but also an important milestone for qubit conversion. 

The authors would like to acknowledge Y. Toda for fruitful discussions.
This work was financially supported by a Grant-in-Aid for Scientific Research from the Ministry of Education, Culture, Sports, Science, Japan. 


\end{document}